# Grade Encoding and the Structural Representation of Student Academic Trajectories

## ——An Empirical Study Based on Encoding Transformation Simulation

**Li Li[1],Yu Cao[2]**

[1]Hefei No.62 Middle School, Hefei 230031,Anhui, China
[2]Anhui NARI ZT Electric Co., Ltd., Hefei 230031, Anhui,China

**Abstract**

Does the conversion of academic assessment from percentage scores to letter grades merely represent an adjustment in information precision, or does it systematically alter the underlying structure of student academic data? Drawing on 68 mathematics exam scores from 75 primary school students, this study employs Encoding Transformation simulation to compare structural differences in the same dataset under two encoding schemes across three dimensions: information loss, distance structure change, and clustering stability. Results indicate that letter-grade encoding compresses the mean pairwise distance in the trajectory feature space from 20.50 to 1.06 (a compression ratio of approximately 19:1); after standardization, the density gradient of the distance distribution is systematically flattened, with kurtosis decreasing by 0.54 and the coefficient of variation decreasing by 0.16;; and the clustering structure becomes highly sensitive to minor sample perturbations—removing a single extreme observation causes the optimal K to jump from 4 to 8 and clustering reproducibility to plummet from 95% to 62%. Concurrently, a Compression Paradox is observed: letter-grade encoding increases the Silhouette coefficient while simultaneously reducing clustering stability. These findings demonstrate that letter-grade encoding systematically alters the structural representation of longitudinal academic data and exerts measurable effects on clustering stability.

## 1.Introduction

### *1.1 Problem Statement*

Since the implementation of the "Double Reduction" policy, academic assessment in China's compulsory education has undergone a structural shift from percentage scores to letter grades[1]. Existing research has largely focused on its socially protective functions, such as alleviating test anxiety[2–4], while rarely interrogating a more fundamental question: Does the change in assessment format merely alter data precision, or does it alter the underlying structure of the data itself?

The significance of this question lies in the following: if encoding schemes can alter the distance structure of data, then all analytical methods that rely on distance computations—cluster profiling, risk early warning, academic portrait construction—will have the stability of their results contingent on the encoding scheme itself. This is not a normative question about "whether the assessment system is good or bad," but an empirical question about "how data structure is shaped by encoding schemes."

*1.2 Research Questions*

The central question of this paper is: After recoding continuous percentage scores into discrete letter grades, does the data structure of student academic trajectories undergo systematic change? If so, what is the mechanism underlying this change?

This is decomposed into three sub-questions:

- RQ1 (Information Layer): How much information loss does letter-grade encoding incur? How does the distance structure of the trajectory feature space change?
- RQ2 (Structure Layer): Is the clustering structure stable after letter-grade encoding? Are clustering results sensitive to sample perturbations?
- RQ3 (Mechanism Layer): If the structure is unstable, what is the generative mechanism?

*1.3 Research Approach*

This study does not rely on a control group in a real policy experiment—under conditions where letter-grade assessment is already widely implemented, such a comparison is institutionally unavailable. Instead, this study employs Encoding Transformation simulation: retaining the original percentage scores of the same cohort of students, converting them to letter-grade data according to current policy rules, and independently executing an identical analytical pipeline under both encoding schemes. This approach logically isolates the net effect of changes in encoding scheme.

*1.4 Preview of Core Findings*

- Letter-grade encoding compresses the mean pairwise distance in the trajectory feature space from 20.50 to 1.06 (compression ratio ≈ 19:1), with nearly 70% information entropy loss[5]
- After standardization, the density gradient of the distance distribution is flattened (kurtosis ↓0.54, CV ↓0.16), and clustering boundaries become highly sensitive to minor perturbations
- Removing a single extreme observation causes the optimal K to jump from 4 to 8, and clustering reproducibility to plummet from 95% to 62%
- Letter-grade encoding exhibits a Compression Paradox: while the Silhouette coefficient[6] increases, clustering stability (PCR/TCR) declines. This directional divergence is termed the Compression Paradox—the two are not contradictory, but are differentiated manifestations of the same compression mechanism at different stages of the data pipeline

## 2. Literature Review

*2.1 Assessment Reform Research: Attention to Function, Neglect of Structure*

In 2020, the Central Committee of the Communist Party of China and the State Council issued the Overall Plan for Deepening Education Assessment Reform in the New Era, explicitly calling for "improving outcome assessment and strengthening process assessment"[7], thereby providing an institutional framework for the shift from percentage scores to letter grades in academic assessment. Existing research has centered on the socially protective functions of this transition. For example, empirical studies have shown that students' and parents' anxiety about academic performance declined significantly after the implementation of the "Double Reduction" policy[3–4]; other research has examined the heterogeneous effects of mathematics anxiety on student academic achievement under the "Double Reduction" context, finding that factors such as test anxiety significantly and negatively affect academic achievement, and proposing corresponding mitigation strategies[2].

The above studies treat the assessment system as a functional instrument, tacitly assuming that the difference between letter grades and percentage scores is limited to precision. However, this presupposition—that changes in data encoding affect only information precision and not the underlying data structure—has yet to be tested as an empirical proposition.

*2.2 Learning Analytics Research: Dependent on Structure, Yet Untested*

Learning analytics transforms academic data from static outcomes into dynamic process signals, relying on distance computations for trajectory modeling, cognitive diagnosis, and precision intervention[8–10]. Yet the field treats academic data as a stable input space—whether distance structure, neighborhood relationships, and topological properties might be altered at the encoding level has not entered the analytical framework[10–11]. Structural stability here serves as a methodological precondition, not as an object of investigation.

*2.3 Educational AI Research: Similarly Dependent on Data Structure*

AI in education research resembles learning analytics in that model performance is highly dependent on data granularity[12–13]. However, this field treats data granularity as a given technical parameter rather than a variable that may be affected by encoding schemes. When high-granularity scores are institutionally withdrawn under policy mandates, the data foundation on which such research depends faces institutional constraints.

*2.4 Common Blind Spot and Research Gap*

In statistics and psychometrics, the cost of artificially discretizing continuous variables has been extensively discussed. Cohen (1983) demonstrated that dichotomization can lead to a loss of up to 55% of explainable variance in the original data[14]; MacCallum et al. (2002) systematically revealed the attenuation of effect sizes and the emergence of spurious nonlinear relationships caused by discretization[11]; Altman and Royston (2006) further noted that the statistical cost paid for this "simplicity" far exceeds researchers' expectations[15]. However, these methodological warnings have largely remained within the domains of statistics and psychology and have yet to enter the research horizon of educational assessment and learning analytics.

The three research pathways above share an untested premise: that the underlying structure of academic data remains stable across changes in encoding scheme. Existing research treats encoding schemes as transparent conduits, not as independent variables capable of altering data structure. The research gap addressed by this paper lies precisely here: treating the encoding scheme itself as an object of study, examining whether and how it alters the distance structure and clustering stability of academic data.

## 3. Analytical Framework

### *3.1 Core Logic*

The analytical framework of this paper is built on a progressive chain:

Encoding Scheme → Information Structure → Distance Structure → Clustering Stability

The encoding scheme determines the number of distinguishable states in the data (information-theoretic basis)[5]; the reduction in distinguishable states alters the distance relationships among samples (distance structure); changes in distance structure affect the partitioning stability of clustering algorithms (clustering stability). These three constitute a complete transmission chain from data input to analytical output.

### *3.2 Three Analytical Layers*

**Table 3-1: Three Analytical Layers**

| Layer | Core Question | Corresponding Indicators | Corresponding Chapter |
|---|---|---|---|
| Information Layer (RQ1) | What does encoding change? | Information entropy, PCA explained variance ratio, pairwise distance distribution | Chapter 5 |
| Structure Layer (RQ2) | Is the clustering structure stable? | Optimal K, PCR, TCR, extreme-value sensitivity | Chapter 6 |
| Mechanism Layer (RQ3) | Why is it unstable? | Magnitude of distance compression, standardization paradox, density gradient, Silhouette–stability divergence | Chapter 7 |

### *3.3 Methodological Positioning of Encoding Transformation Simulation*

The Encoding Transformation simulation employed in this study differs from the counterfactual framework in causal inference. This study does not claim to estimate "the causal effect of letter-grade assessment"—such an effect is unidentifiable in the real world, as no single education system can simultaneously accept two encoding schemes. The logic of this study is a more fundamental data-level comparison: for the same students and the same exams, changing only the numerical encoding scheme, observing whether the data structure changes. This design isolates the net effect of the encoding scheme, not the causal effect of policy implementation.

## 4. Data and Encoding Transformation

### *4.1 Data Source*

The data come from longitudinal mathematics achievement records of two classes in a primary school spanning Grades 1 through 5, comprising 75 students and a total of 68 exams. All scores were originally recorded as percentage scores. Data collection was conducted with school and guardian consent, and all personal information has been anonymized.

Mathematics was selected based on the following considerations: first, mathematics exam scoring is relatively highly standardized, with scores less susceptible to subjective interference; second, the cumulative and hierarchical nature of mathematical knowledge enables the construction of longitudinal academic development trajectories; third, mathematics achievement during primary school is highly representative of academic performance across compulsory education.

*4.2 Grade Conversion Rules*

In accordance with the current letter-grade assessment rules of Menghai County[16], percentage scores are converted into four letter grades: A/B/C/D:

**Table 4-1: Grade Conversion Rules**

| Original Score | Grade | Numerical Encoding |
|---|---|---|
| ≥90 | A | 4 |
| 80–89 | B | 3 |
| 60–79 | C | 2 |
| <60 | D | 1 |

Both encoding schemes share the same cohort of students and their original score data; the sole difference lies in the numerical encoding form of the data.

*4.3 Trajectory Feature Extraction*

For each student's 68 exam scores, six trajectory features are extracted:

**Table 4-2: Trajectory Feature Extraction**

| Feature | Definition | Meaning |
|---|---|---|
| Mean | Average score across 68 exams | Academic level |
| Standard Deviation | Standard deviation of 68 exam scores | Fluctuation amplitude |
| Trend Slope | Slope of linear regression of scores against exam sequence | Long-term developmental trend |
| Range | Maximum score minus minimum score | Score span |
| Maximum Score | Highest score across 68 exams | Peak performance |
| Rate of Change | Standard deviation of successive differences | Short-term fluctuation intensity |

All six features are standardized (z-score) before entering PCA dimensionality reduction and K-means clustering.

### *4.4 Analytical Pipeline*

An identical analytical pipeline is executed under both encoding schemes to ensure that any observed differences arise solely from the encoding scheme itself:

1. Extract six-dimensional trajectory features
2. Standardize (z-score)
3. PCA dimensionality reduction
4. K-means clustering (K = 2–8 sweep, Silhouette coefficient selects optimal K)
5. Clustering reproducibility test (Leave-one-out PCR)
6. Cross-window stability test (cumulative-window TCR)

## 5. Data Structure Changes Induced by Letter-Grade Encoding

Core question: What changed? (RQ1)

### *5.1 Information Entropy Loss*

Information entropy (Shannon Entropy) measures the number of distinguishable states in the data[5]. The information entropy of percentage scores is 5.16 bits (86 distinguishable states), while that of letter grades is 1.60 bits (4 distinguishable states), yielding an information entropy loss rate of 69.01%. Letter-grade encoding compresses score values from 86 distinguishable states into 4, with nearly 70% of information-carrying capacity systematically stripped away.

### *5.2 Decline in PCA Explained Variance*

PCA dimensionality reduction is applied to the six-dimensional trajectory features. The explained variance ratios under the two encoding schemes are compared below:

**Table 5-1: PCA Explained Variance Comparison**

| Data Form | PC1 Explained Variance | PC2 Explained Variance | Cumulative Explained Variance |
|---|---|---|---|
| Percentage | 77.62% | 13.28% | 90.90% |
| Letter Grade | 59.00% | 20.47% | 79.47% |

Under percentage scores, the first two principal components summarize 90.90% of the variance; under letter grades, only 79.47%—a decline of over 11 percentage points. Changes in encoding scheme not only reduce the amount of information but also diminish the capacity of linear dimensionality reduction to summarize data structure—in the letter-grade-encoded trajectory feature space, more variance is dispersed into high-dimensional noise.

### *5.3 Pairwise Distance Compression*

In the unstandardized, raw six-dimensional trajectory feature space, the distributions of pairwise Euclidean distances among the 75 students under the two encoding schemes are compared:

**Table 5-2: Pairwise Distance Compression Comparison (Raw Scale)**

| Statistic | Percentage | Letter Grade |
|---|---|---|
| Mean | 20.50 | 1.06 |
| Std. Dev. | 17.03 | 0.71 |
| Maximum | 97.56 | 3.23 |

Under letter-grade encoding, the mean pairwise distance drops from 20.50 to 1.06, a compression ratio of approximately 19:1. Using the P5 threshold of the percentage-score distance distribution as a benchmark, 99.75% of student-pair distances under letter-grade encoding fall within that threshold—all students are squeezed into an extremely dense distance space. The "internal flattening" effect at the raw score level—within-grade differences being zeroed out—after aggregation across 68 exams, is not canceled out through averaging but is cumulatively transmitted to the trajectory feature space in the form of an overall contraction of the distance field.

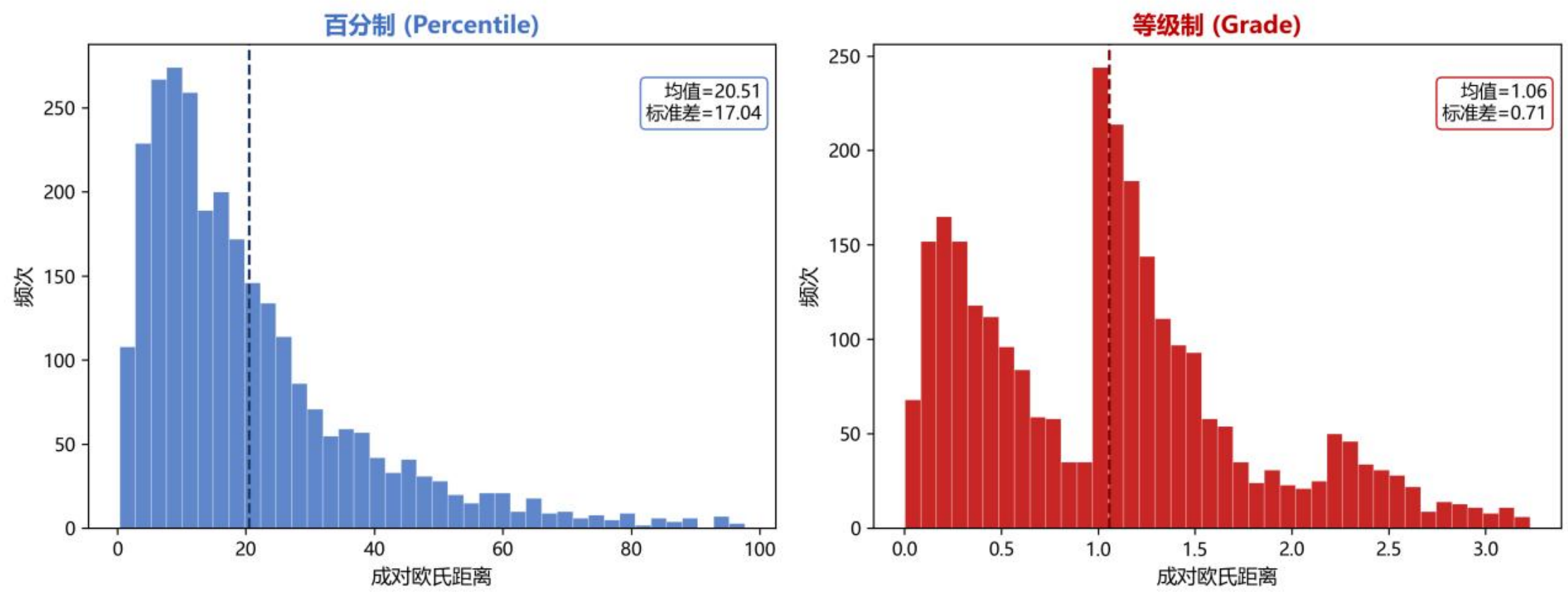

**Figure 5-1: Comparison of pairwise distance distributions on the raw scale.**

Left and right subplots show histograms of pairwise Euclidean distances among 75 students under percentage scores (blue) and letter grades (red), respectively. The percentage-score distance distribution spans a wide range, with a long tail extending to nearly 100 and a mean of 20.50; the letter-grade distance distribution is compressed into an extremely narrow interval of 0–3, with the mean plummeting to 1.06—a difference of approximately 19-fold in magnitude.

*5.4 Skewness Reversal and Dimensional Degeneration in Feature Distributions*

Letter-grade encoding not only compresses distances but also leaves discernible micro-level imprints on the underlying feature distributions. Skewness and kurtosis are computed for each of the six trajectory features:

**Table 5-3: Comparison of Trajectory Feature Distribution Shapes**

| Feature | Percentage Skewness | Letter-Grade Skewness | Skewness Change | Key Signal |
|---|---|---|---|---|

| Mean | −2.63 | −1.42 | +1.21 | Sign preserved |
|---|---|---|---|---|
| Std. Dev. | +1.45 | −0.64 | −2.08 | Sign reversal |
| Trend Slope | −0.68 | −0.12 | +0.56 | Sign preserved |
| Range | +1.06 | −0.07 | −1.13 | Sign reversal |
| Maximum Score | −3.56 | −8.49 | −4.93 | Near-constant degeneration |
| Rate of Change | +1.38 | −0.54 | −1.92 | Sign reversal |

Two salient signals emerge. First, skewness sign reversal: the skewness of standard deviation and rate of change flips from positive to negative—the grade-boundary amplification effect causes more students' trajectories to exhibit elevated fluctuation levels, systematically reshaping the distribution shape. Second, dimensional degeneration: under letter-grade encoding, the maximum score feature takes the value 4 for the vast majority of students (Grade A achieved at least once), with skewness reaching −8.49 and kurtosis as high as 70.01—this dimension essentially degenerates into a near-constant, leaving only five effective dimensions in what was once a six-dimensional feature space.

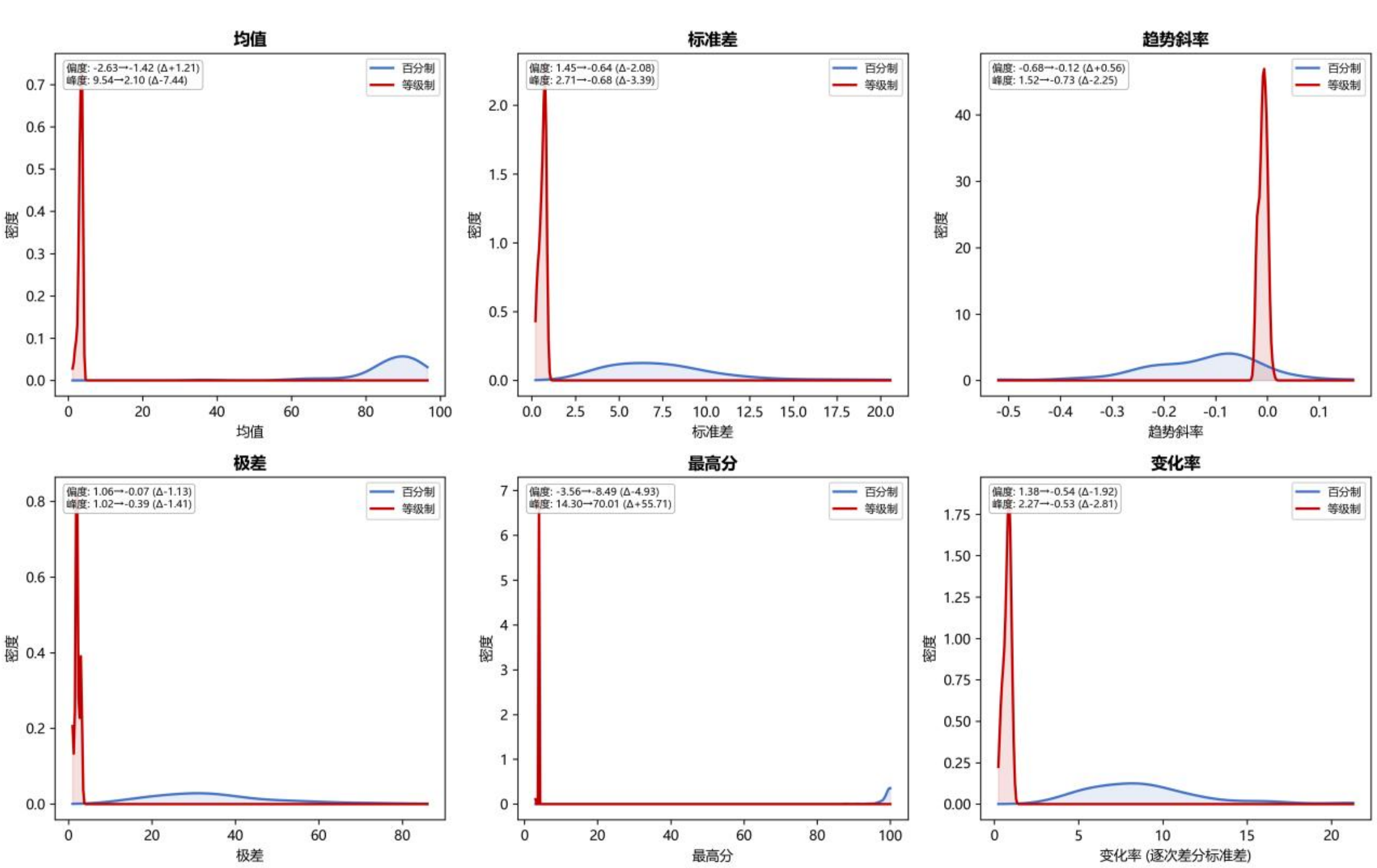

**Figure 5-2: Comparison of probability density distributions of trajectory features**

Six subplots respectively display the density distributions of the six trajectory features under percentage scores (blue KDE curves) and letter grades (red KDE curves), with skewness and kurtosis change values annotated in each subplot. Skewness sign reversal (positive → negative) is visible in the standard deviation and rate of change subplots for the red distribution; the maximum score subplot shows the red distribution collapsing to a single point (near-constant degeneration).

### *5.5 Summary*

Letter-grade encoding systematically alters data structure across three dimensions: a 69% loss in information entropy, a decline of over 11 percentage points in cumulative PCA explained variance, and an approximately 19:1 compression of pairwise distances. Furthermore, discretization leaves micro-level imprints of skewness reversal and dimensional degeneration in the feature distributions. These changes constitute the data foundation for the subsequent clustering stability analysis.

## 6. Clustering Stability

Core question: What happened to clustering? (RQ2)

### *6.1 Full-Sample Structural Profiling (n = 75)*

Under full-sample conditions, K-means clustering is conducted for K = 2 through 8, with the Silhouette coefficient used to select the optimal K.

**Table 6-1: Silhouette Coefficient Sweep (Full Sample, K = 2–8)**

| K | Percentage Silhouette | Letter-Grade Silhouette |
|---|---|---|
| 2 | 0.60 | 0.52 |
| 3 | 0.43 | 0.54 |
| 4 | 0.43 | 0.55 |
| 5 | 0.40 | 0.55 |
| 6 | 0.40 | 0.54 |
| 7 | 0.40 | 0.50 |
| 8 | 0.42 | 0.50 |

Under percentage scores, the optimal K = 2 (Silhouette = 0.60); under letter grades, the optimal K = 4 (Silhouette = 0.55). Percentage scores exhibit a clear unimodal silhouette peak at K = 2, whereas under letter grades the silhouette differences among K = 3–5 are negligible (0.54–0.55), lacking a clear global optimum—this itself signals an indistinct density gradient in the partitioning structure.

### *6.2 Extreme-Value Sensitivity Analysis*

To test the sensitivity of the clustering structure to extreme observations, the observation with the most extreme PC1 score (PC1 = 9.82 under percentage scores, markedly deviating from the main distribution) is removed, and the entire analytical pipeline is re-executed (n = 74).

**Table 6-2: Extreme-Value Sensitivity Analysis (n = 75 vs. n = 74)**

| Indicator | n = 75 | n = 74 |
|---|---|---|
| Letter-Grade Optimal K | 4 | 8 |
| Leave-One-Out PCR | 95.25% | 61.58% |
| Percentage Optimal K | 2 | 2 |
| Percentage PCR | 97.76% | 93.24% |

After removing a single observation, the letter-grade optimal K jumps from 4 to 8, and the Leave-One-Out clustering reproducibility (PCR) plummets from 95% to 62%—the removal of a single observation triggers a wholesale reconfiguration of clustering boundaries. Under percentage scores, the same operation leaves the optimal K unchanged at 2, with PCR only modestly declining from 98% to 93%. The differential sensitivity to extreme values between the two encoding schemes indicates that the clustering structure under letter-grade encoding is highly dependent on sample composition; the K = 4 observed under full-sample conditions is not a robust data characteristic, but a temporary equilibrium under a specific sample configuration.

*6.3 Cross-Window Consistency (TCR)*

The first 20, 30, 40, 50, 60, and all 68 exam records are respectively extracted to form six cumulative time windows. Each window independently undergoes the complete feature extraction → standardization → PCA → K-means clustering pipeline, retaining its own optimal K. The Adjusted Rand Index (ARI) between each window's cluster labels and those of the final window (W68) is computed as the cross-window consistency indicator (TCR).

**Table 6-3: Cross-Window Consistency (TCR)**

| Window Comparison | Percentage TCR | Letter-Grade TCR |
|---|---|---|
| W20 vs. W68 | 0.93 | 0.52 |
| W30 vs. W68 | 0.96 | 0.68 |
| W40 vs. W68 | 0.94 | 0.78 |
| W50 vs. W68 | 0.95 | 0.85 |
| W60 vs. W68 | 0.93 | 0.96 |

Under percentage scores, TCR remains consistently high at 0.93–0.96; even when using only the first 20 exams (less than one-third of the total data), the cluster labels are already highly consistent with the final window—the students' academic structure itself is highly stable over time, and the accumulation of additional information does not alter the fundamental profiling. This finding also constitutes an important counterfactual exclusion: if the difference in TCR between percentage and letter-grade encoding stemmed from "genuine academic change," the percentage TCR should likewise exhibit substantial fluctuation; the persistently high percentage TCR pins the source of the TCR difference squarely on the encoding scheme itself.

Under letter-grade encoding, TCR climbs slowly from 0.52 to 0.96. The cluster labels from the first 40 exams (W40) show only 0.78 consistency with the final window—if students were clustered under this window, nearly a quarter of labels would change after subsequent data accumulation. Structural convergence is far slower than under percentage scores, and the convergence path is highly dependent on the number of exams.

*6.4 Summary*

The clustering structure under letter-grade encoding exhibits conditional fragility: superficially stable under full-sample conditions (K = 4, PCR ≈ 95%), yet the K value jumps and PCR plummets upon removal of a single extreme observation (§6.2), and cluster labels drift substantially when the time window is altered (§6.3). These instabilities do not arise from genuine changes in students' academic structure—percentage scores remain consistently stable—but from changes in the underlying properties of the data structure induced by letter-grade encoding. The next chapter analyzes the generative mechanism of this alteration.

## 7. Mechanism Analysis — Distance Compression and the Compression Paradox

Core question: Why did it happen? (RQ3)

*7.1 Transmission Chain (Detailed)*

The effect of letter-grade encoding on clustering stability is not a direct linear transmission but operates through three stages—compression, standardization, and density flattening—with the core mechanism as follows:

- Stage 1: Compression. Letter-grade encoding zeroes out score differences within the same grade (the "internal flattening" effect). After cumulative transmission across 68 exams, the mean pairwise distance in the trajectory feature space plummets from 20.50 to 1.06 (compression ratio ≈ 19:1), squeezing all students into an extremely dense distance space.
- Stage 2: Standardization Paradox. $z$ -score standardization eliminates magnitude differences, yet forcibly stretches residual tiny relative differences in the compressed space (e.g., the boundary fluctuation between scores of 89 and 90) to the same magnitude as genuine differences. This causes the kurtosis of the distance distribution to drop by 0.54 and the coefficient of variation (CV) to drop by 0.16, systematically shaving off the sharp density peaks.
- Stage 3: Density Flattening and Clustering Destabilization. Once density peaks and valleys are shaved off, the spatial topological form tends toward a gradient-free "plain," and the optimal position of the algorithm's separating hyperplane is no longer unique. This renders clustering boundaries highly sensitive to minor perturbations, manifesting as K-value jumps under sample perturbation (PCR dropping to 61.58%) and label drift as the time window advances (TCR as low as 0.52).

The pivotal link in this chain is the standardization paradox, which is analyzed below.

*7.2 The Standardization Paradox*

PCA and K-means clustering require feature standardization. After standardization, the distance means under the two encoding schemes converge (percentage: 2.77 vs. letter grade: 2.99), but the distribution shapes undergo qualitative changes:

**Table 7-1: Comparison of Standardized Pairwise Distance Distribution Shapes**

| Statistic | Percentage | Letter Grade | Change |
|---|---|---|---|
| Skewness | 1.83 | 1.52 | ↓0.31 |
| Kurtosis | 4.01 | 3.47 | ↓0.54 |
| Coefficient of Variation (CV) | 0.76 | 0.60 | ↓0.16 |

Standardization, while eliminating magnitude differences, stretches tiny relative differences in the compressed space to the same magnitude as genuine differences, creating extensive regions of pseudo-heterogeneity. Both the kurtosis and CV of the distance distribution decline—under percentage scores, the distance distribution forms a sharp density peak near the mean, whereas under letter-grade encoding the density peak is systematically shaved off. The direct consequence is that the density peaks on which clustering algorithms rely to identify natural groupings tend to vanish, and the optimal position of the separating hyperplane is no longer unique.

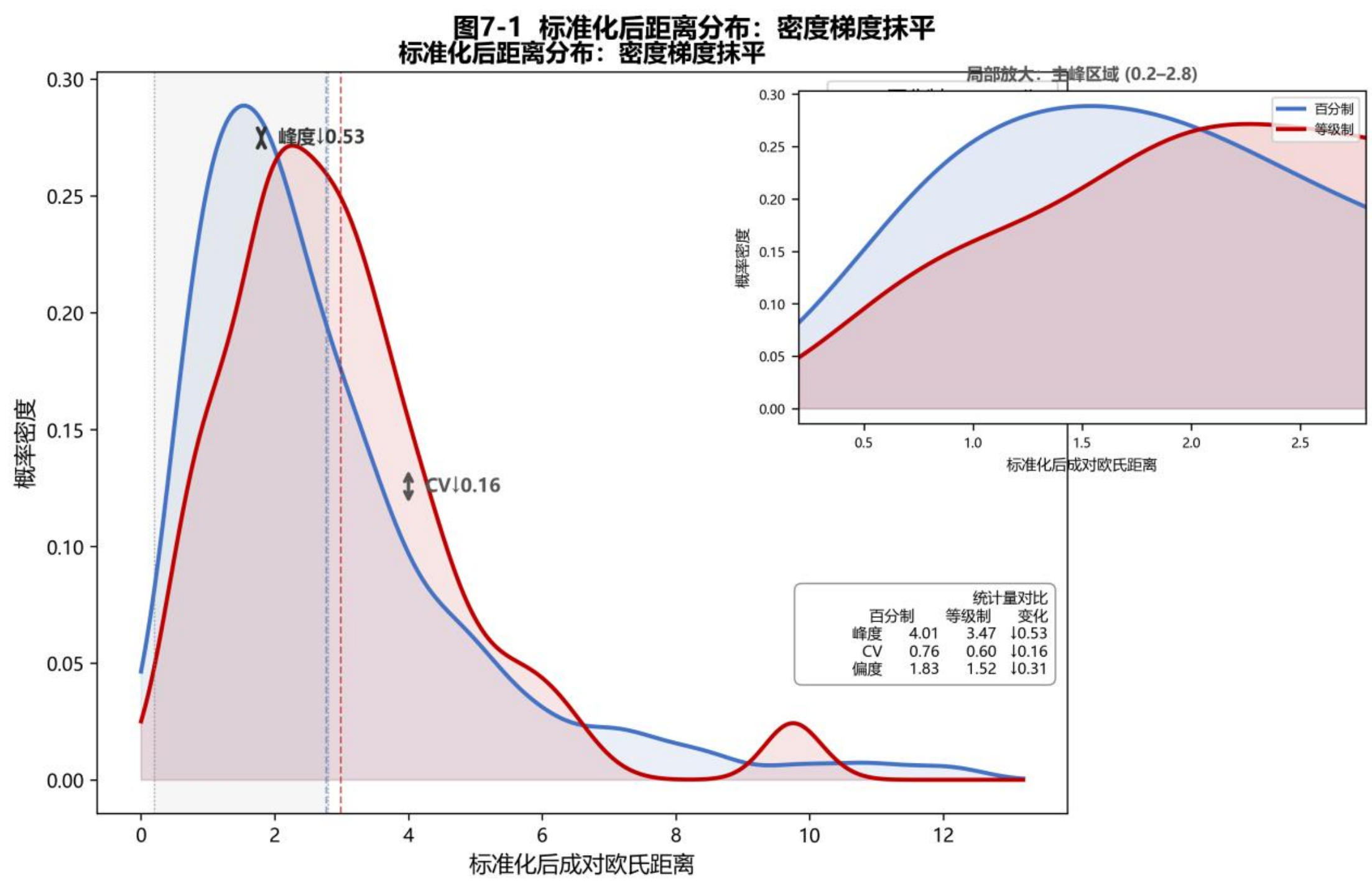

**Figure 7-1: Standardized distance distribution — density gradient flattening.**

The main plot overlays the KDE density curves of standardized pairwise distances under both encoding schemes: the blue curve (percentage) exhibits a sharp, towering main peak, while the red curve (letter grade) shows a markedly flatter main peak with wider tails. Bidirectional arrows annotate the kurtosis reduction (↓0.54) and CV reduction (↓0.16). The lower-right inset zoom window focuses on the main peak region (x: 0.2–2.8), further illustrating the structural difference between the concentrated density peak under percentage scores and the diffuse density form under letter-grade encoding.

*7.3 The Compression Paradox*

Letter-grade encoding produces directionally opposite effects at different stages of the data pipeline, constituting what this paper terms the Compression Paradox:

**Table 7-2: Stage-Differentiated Effects of the Compression Paradox**

| Stage | Data State | Effect | Indicator Manifestation |
|---|---|---|---|
| Before Standardization | Raw scale | Discretization forcibly draws same-grade students closer; within-cluster variance artificially reduced | Silhouette ↑ (percentage 0.43 → letter grade 0.55) |
| After Standardization | Standardized feature space | Density gradient flattened; separating hyperplane not unique | PCR ↓ (95% → 62%), TCR ↓ (0.93 → 0.52) |

The two are not contradictory—the Silhouette increase is the static appearance of compression, while the PCR/TCR decline is the dynamic consequence of structural destabilization. The same compression mechanism produces directionally opposite indicator manifestations at two stages of the data pipeline.

This finding carries methodological cautionary implications for educational data analysis practice. The field of educational data mining widely relies on the Silhouette coefficient to select the number of clusters and evaluate clustering quality [6, 10]. However, the results of this study indicate that under letter-grade encoding conditions, a higher Silhouette does not correspond to a more stable clustering structure—quite the contrary, the K = 4 partition "endorsed" by the Silhouette coefficient collapses upon removal of a single observation. Under conditions where the encoding scheme may alter the underlying data structure, the Silhouette coefficient alone cannot serve as a reliable criterion for clustering quality.

*7.4 Phenomenon Designation: Data Silence*

This paper designates the phenomenon—whereby changes in encoding scheme lead to the compression of data structure and the decline of clustering stability—as Data Silence.

Data Silence is distinct from "Data Missingness." Data Missingness refers to the physical scarcity of data—unmeasured, uncollected, or lost—which can be remedied by increasing collection efforts. Data Silence refers to a situation in which data physically exist (the original percentage scores have not disappeared), yet changes in encoding scheme systematically constrain the expression of differences within the data structure—distance compression, density gradient flattening, and clustering highly sensitive to minor perturbations. The damage lies not in the reduction of data quantity, but in the deformation of data structure. This phenomenon suggests that simply increasing sample size may not restore the data structure changes caused by encoding; the scope of its applicability awaits further research.

## 8. Discussion and Conclusion

*8.1 Main Findings*

1. Encoding schemes alter data structure. After recoding the same set of percentage scores into letter grades, the distance structure of the trajectory feature space undergoes systematic change—the mean pairwise distance drops from 20.50 to 1.06 (compression ratio ≈ 19:1), and the density gradient is flattened. Data encoding schemes are not transparent conduits.
2. Clustering stability is highly sensitive to encoding schemes. The clustering structure under letter-grade encoding exhibits conditional fragility—superficially stable under full-sample conditions (K = 4), yet after removal of an extreme value, K jumps to 8 and PCR plummets to 62%; after changes in the time window, TCR drops as low as 0.52.
3. The existence of the Compression Paradox. Letter-grade encoding simultaneously elevates the Silhouette coefficient and reduces clustering stability; under conditions of encoding scheme change, the Silhouette coefficient alone cannot serve as a reliable criterion for clustering quality.

*8.2 Academic Contributions*

1. Provides empirical evidence that encoding schemes affect data structure and clustering stability. Existing research treats encoding schemes as transparent preconditions; this study finds that the encoding scheme itself is an independent variable affecting clustering outcomes.
2. Discovers and designates the Compression Paradox. The directional divergence between the Silhouette coefficient and stability indicators under letter-grade encoding constitutes a substantive caution against the routine practice of using the Silhouette coefficient for cluster number selection under conditions of encoding scheme change.
3. Proposes "Data Silence" as a designation for the phenomenon of encoding effects. Distinct from Data Missingness, Data Silence points to the constrained structural expression and decline in clustering stability caused by encoding scheme changes; the damage lies in structural deformation rather than the reduction of information quantity.

*8.3 Practical Implications*

The findings of this paper do not constitute a normative judgment on letter-grade assessment policy. This paper merely points out that when the education system adopts letter grades as the primary presentation format for academic data, the stability of distance-based analytical methods—such as cluster profiling, academic portrait construction, and risk early warning—applied to such data will face encoding-level uncertainty. How to preserve analytical space for high-granularity data for educational diagnosis while maintaining the burden-reducing function of letter-grade assessment merits further exploration.

*8.4 Limitations and Future Directions*

1. Sample limitations. The 75 students come from a single school and a single subject (mathematics). What this study reveals is the existence of encoding effects, not their overall effect size. Multi-region, multi-subject, large-sample validation is a necessary follow-up endeavor.

2. Algorithm limitations. This study employs K-means as a detector of distance structure changes. The mechanism analysis of this study indicates that discretization alters the underlying distance structure of the input data—this structural deformation lies upstream of all distance-based algorithms. The performance of other algorithms such as GMM, DBSCAN, and spectral clustering under letter-grade encoding conditions merits systematic comparison.
3. Encoding rule limitations. This study adopts a four-tier system (A/B/C/D). The effects of different grading schemes (three-tier, five-tier, equal-interval/unequal-interval cutoffs) may differ.
4. Boundaries of encoding transformation simulation. The Encoding Transformation simulation employed in this study isolates the net effect of the encoding scheme. In real policy implementation, teacher grading behavior may undergo strategic adjustments in response to changes in the assessment system (e.g., assigning "safe scores" near grade boundaries); this behavioral adaptation effect lies outside the analytical scope of this study.